\documentstyle[12pt,epsf]{article}
\newlength{\dinwidth}
\newlength{\dinmargin}
\begin{document}

\def\thefootnote{\fnsymbol{footnote}}

\baselineskip18pt

\thispagestyle{empty}

\begin{flushright}
\begin{tabular}{l}
  FTUAM-93/14\\\vspace*{24pt} May, 1993
\end{tabular}
\end{flushright}

\vspace*{1.5cm}

{\vbox{\centerline{{\Large{\bf STRING VARIATIONS}}}}}

\vskip12pt

{\vbox{\centerline{{\Large{\bf ON KALUZA-KLEIN COSMOLOGY}}}}}

\vskip72pt\centerline{M.\,A.\,R. Osorio\footnote{E-mail addresses:
    {\tt OSORIO@VM1.SDI.UAM.ES} and {\tt
      OSORIO@MADRIZ1.FT.UAM.ES}.}\footnote{Address after June 1st.
    1993, Departamento de F\'{\i}sica, Facultad de Ciencias,
    Universidad de \newline\indent\hspace{5pt} Oviedo, Avda. Calvo
    Sotelo s/n, E-33007 Oviedo, Spain.  } and M.\,A.
  V\'azquez-Mozo\footnote{E-mail addresses: {\tt
MAVAZ@VM1.SDI.UAM.ES}
    and {\tt VAZQUEZ@MADRIZ1.FT.UAM.ES}.}}

\vskip12pt
\centerline{{\it Departamento de F\'{\i}sica Te\'orica
C-XI}}\vskip2pt
\centerline{{\it Universidad Aut\'onoma de Madrid}}\vskip2pt
\centerline{{\it 28049 Madrid, Spain}}

\vskip .7in

\baselineskip24pt
\indent

We study the cosmological solutions of the two-dimensional
Brans-Dicke
equations considering a gas of $c=1$ strings in $S^{1}\times
\mbox{\bf
  R}$ as the source of the gravitational field. We also study the
implications of the $R$-duality invariance on the solutions. To this
purpose we conjecture that, as it happens for massless fields in
finite boxes, the free energy of a gas of massless string excitations
is not given by the corresponding toroidal compactification.

\baselineskip18pt

\setcounter{page}{0}

\def\thefootnote{\arabic{footnote}}

\setcounter{footnote}{0}

\newpage

\section{Introduction}

In a previous paper \cite{OV-M-1} we have studied the implications of
the presence of compactified spatial dimensions (or equivalently, the
non-trivial topology of the space-time) on the cosmological solutions
of a two-dimensional toy model coupled to Brans-Dicke gravity.  There
we found that the only observable effect of the closed dimension
appears in the thermal history.  The universe with compactified
dimensions gets cold more slowly than in the open space case.  The
dynamical evolution, on the contrary, remains exactly the same.  The
reason for this is that the introduction of non-trivial topology in
the space does not modify the equation of state associated with the
(field-theoretical) matter filling the universe.

Now we would like to study the cosmological solutions to the
Brans-Dicke equations when the source of the gravitational field is a
gas of strings living in a two-dimensional target space. From the
very
beginning we face a problem which is that of determining the correct
expression of the free energy. If we take as true the conjecture of
\cite{TV} in which the free energy is naively identified with the
toroidal compactification in $S^{1}\times S^{1 }$, the only remnant
question to clarify is that of identifying the time coordinate. If we
restrict our interest up to one loop in the world-sheet the question
is irrelevant since the space dimension is compact and $R^{(2)}=0$,
then this is only a problem at higher genus ($g\geq 2$). If we took
the Liouville field as the time coordinate its coupling to the
world-sheet curvature would imply that only discrete values of the
temperature are allowed \cite{TV}. On the contrary, based on
aesthetical grounds, it looks more appealing to us to take the
Liouville field as a spatial coordinate and then the length of the
target coordinate of the $c=1$ string is $\beta$. This leads to a
picture in which the string interaction produces a space whose length
is quantized.

Following the conjecture of \cite{TV}, an expression for the
Helmholtz
free energy is gotten which reproduces the partition function of the
$c=1$ non-critical string on a two-dimensional target torus
\cite{DKL}
(the concrete relation is that the sum over world-sheet surfaces
equals $-\beta F(\beta)$)
\begin{equation}
  F(\beta,L)=\frac{1}{\beta}\ln{\left(\frac{L}{\sqrt{\alpha^{'}}}
  \right)} + \frac{2}{\beta}\ln{\left[\eta\left(i \frac{\beta
    L}{4\pi^{2}\alpha^{'}}\right)
  \eta\left(i\frac{L}{\beta}\right)\right]}\:,\label{free-energy}
\end{equation}

where $L$ is the length of the spatial compactified dimension and
$\eta(\tau)$ is the Dedekind $\eta$-function.  The equation of state
is
\begin{equation}
  \rho-p=\frac{1}{\beta L}-\frac{1}{12\pi\alpha^{'}}
  E_{2}\left(i\frac{\beta L}{4\pi^{2}\alpha^{'}}\right)\;,
\end{equation}
with $E_{2}(\tau)$ the Eisenstein series.  In this case we have some
extra symmetries that are absent in the field-theoretical case,
namely, the thermal partition function $Z(\beta,L)$ enjoys both
$\beta$ and space-time duality
\begin{equation}
  Z(\beta,L)=Z\left(\frac{4\pi^{2}\alpha^{'}}{\beta},L\right)=
  Z\left(\beta,\frac{4\pi^{2}\alpha^{'}}{L}\right)
\end{equation}
Since the Einstein-Hilbert-Brans-Dicke action
\begin{equation}
  S=\int d^{2}x \sqrt{-g}\left[\Phi(R-2\Lambda)-\frac{\omega}{\Phi}
  \nabla_{\mu}\Phi \nabla^{\mu}\Phi\right]
\label{EHBD}
\end{equation}
with the Friedmann-Robertson-Walker {\it ansatz} for the metric
\begin{equation}
  ds^{2}=-dt^{2}+L^{2}(t)d\xi^{2}
\label{FRW}
\end{equation}
and a space-independent Brans-Dicke field $\Phi(t)$ is invariant
under
the duality replacement
\begin{equation}
  L \rightarrow \frac{C}{L}\:, \hspace{2cm} \Phi\rightarrow
  \frac{L^{2}}{C}\Phi\;,
\end{equation}
with $C$ any constant, we see that the whole action ({\it fields +
  matter}) is invariant under the space-time duality transformation
\begin{equation}
  L \rightarrow \frac{4\pi^{2}\alpha^{'}}{L}\;, \hspace{3cm} \Phi
\rightarrow \frac{L^{2}}{4\pi^{2}\alpha^{'}}\Phi\;.
\end{equation}
This is not true for $\beta$-duality. The reason is that only the
matter action depends on the temperature and it transforms
non-trivially under duality. Then the changes in the action of the
matter induced by the duality transformation cannot be undone by the
variation of the Einstein-Hilbert-Brans-Dicke action since this does
not depend on $\beta$. The result is that the equations governing the
dynamics of our universe are invariant under space-time duality but
not under $\beta$-duality.

It seems then that $\beta$-duality does not survive the coupling to
the Einstein-Hilbert-Brans-Dicke action. Furthermore, what is really
a
problem is that (\ref{free-energy}) is positive definite for some
region in the $\beta-L$ plane and actually it suffers from a fatal
disease, the entropy is negative and divergent as $\beta \rightarrow
\infty$. The only possible conclusion is that the partition function
for the $c=1$ model on a torus, given by the integral on the
fundamental region of the modular group of the solitonic
contribution,
does not coincide with minus the logarithm of the thermal partition
function $Z(\beta,L)$.  This is not a surprise because we have found
the same situation for massless free fields in $ S^{1}\times\mbox{\bf
  R}$ \cite{OV-M-1}\footnote{$L$ is the length of $S^{1}$} and, after
all, the $c=1$ model can be effectively described by a massless
scalar
field in two dimensions \cite{P}. It is worth noticing that
(\ref{free-energy}) can be written
\begin{equation}
  F(\beta,L)=
\frac{1}{\beta}\ln{\left(\frac{\beta}{\sqrt{\alpha^{'}}}
  \right)} + F_{B}(\beta,L) +
F_{B}\left(\beta,\frac{4\pi^{2}\alpha^{'}}{L}\right)
\label{new}
\end{equation}
where $F_{B}(\beta,L)=(2/\beta)\ln{\eta(i\beta/L)}$ is the free
energy
of a massless boson in $S^{1}\times \mbox{\bf R}$ \cite{OV-M-1}. Our
proposal is that dropping the first term which depends only on
$\beta$
we get the correct result for the one-loop free energy of the $c=1$
model in $S^{1}\times\mbox{\bf R}$. The price we have to pay is to
renounce to $\beta$-duality. Nevertheless, we cure the {\it maladie}.

The plan of the paper is as follows. In section 2 we will study the
main features of the thermodynamics of the critical two-dimensional
string in $S^{1}\times \mbox{\bf R}$ and its relationship with a
bosonic field living in the same space. In section 3 we will present
the cosmological solutions to the Brans-Dicke equations. Finally in
section 4 we will summarize the conclusions.

\section{The thermodynamics of the string in $S^{1}\times \mbox{\bf
R}$}

The quantization of the bosonic string in arbitrary space-time
dimension is a hard problem that still remains unsolved in general.
The reason is that away from the critical dimension ($d=26$) there is
an anomaly associated with the conformal invariance of the string
action \cite{Polyakov}. Then the conformal factor of the metric (the
Liouville field) becomes a dynamical field that has also to be
quantized. The quantization of the Liouville theory can be made in
the
particular case of the $c=1$ conformal field theory coupled to
two-dimensional gravity \cite{DDK}.  In this case, after identifying
the Liouville field as a new coordinate and with a suitable dilatonic
background, the $c=1$ string can be reinterpreted as a critical
two-dimensional string theory \cite{GM}.

We are going to consider that the matter content of the universe is
given by a gas of these two-dimensional critical strings. If we
compactify the spatial coordinate in a circumference of radius
$R=L/2\pi$, and take (\ref{free-energy}), i.e., the result of the
toroidal compactification as computed in \cite{DKL} (cf.
\cite{OV,TV}), as the free energy we see that our thermodynamics
enjoys both space-time and $\beta-$duality,
\begin{equation}
  F(\beta,L)= F\left(\beta, \frac{4\pi^{2}\alpha^{'}}{L}\right)=
  \frac{4\pi^{2}\alpha^{'}}{\beta^{2}}
  F\left(\frac{4\pi^{2}\alpha^{'}}{\beta},L\right)\;.
\end{equation}
In addition to this, $-\beta F(\beta,L)$ is invariant under the
interchange $\beta \leftrightarrow L$, as it should be if the free
energy is gotten from the toroidal compactification in $S^{1}\times
S^{1}$.

Let us review the main thermodynamical properties that can be
extracted from (\ref{free-energy}). The first important thing to
mention is that $F(\beta,L)$ is not always negative. One can think
that, after all, being the result of a regularization procedure some
arbitrary term of the form $constant/\beta$ can be added. Physically,
adding this term is equivalent to tuning the value of the canonical
entropy at zero temperature ($\beta=\infty$). Then one can try to fix
the constant to get zero entropy at zero temperature. From the very
beginning this is not a legitimate manipulation because in quantum
statistical mechanics the degeneration of the vacuum is what it is.
But, we do not even need to think about this problem because fixing
the entropy at zero temperature to zero by adding that term is
impossible since the entropy behaves as minus $\ln{\beta}$ as $\beta
\rightarrow \infty$ (see below).

A similar phenomenon is observed in critical strings at finite
temperature \cite{AO-PA} where the dual phase has {\it bounded}
negative entropy. Here one is also tempted to add the adequate
constant to set the entropy positive since at least it is only a
finite constant. But one must be careful because standard texts in
thermodynamics and statistical mechanics (see for example
\cite{Balescu}) teach us that the entropy measures the number of
states at a given energy that can be excited. But in String Theory it
seems that at the self-dual point new degrees of freedom are excited
\cite{OV-M2} that kill the low energy field-theoretical ones. Let us
concentrate for a moment in the particular case of the supersymmetric
heterotic string at finite temperature \cite{AW,AO-NP}.
$\beta$-duality implies that at high temperature there are degrees of
freedom which act still as fermions so as to recover a high
temperature version of supersymmetry. In standard statistical
mechanics it is precisely the equivalence between Bose, Fermi and
Maxwell-Boltzmann statistics at high temperature and low density
which
fixes the additive constant in the entropy. Our conclusion is that
all
these things, namely, the statistics of the string itself and the
physical degrees of freedom that it represents, are involved in the
resolution of this problem that is a mystery to us. In the problem we
have now at hand, by inspecting the expressions for the energy and
the
pressure, we can see that high temperature and low density can
correspond either to a stringy or a field-theoretical regime.

The canonical entropy obtained from (\ref{free-energy}) is
\begin{eqnarray}
  S(\beta,L)&=& -\frac{\beta L}{12 \pi \alpha^{'}}
  E_{2}\left(i\frac{\beta L}{4 \pi^2 \alpha^{'}}\right) +\frac{\pi
    L}{6 \beta} E_{2}\left(i\frac{L}{\beta}\right) - \beta F(\beta,L)
  \nonumber \\ &=& 1-\ln{\frac{\beta}{\sqrt{\alpha^{'}}}}+
  S_{B}(\beta,L)+
  S_{B}\left(\beta,\frac{4\pi^{2}\alpha^{'}}{L}\right)\;,
\label{entropy}
\end{eqnarray}
where
\begin{equation}
  S_{B}(\beta,L)=-2
  \ln{\eta\left(i\frac{\beta}{L}\right)}-\frac{\pi\beta}{6L}E_{2}
  \left(i\frac{\beta}{L}\right)
\end{equation}
is the entropy for a massless bosonic field. From this expression we
see that when $\beta$ goes to infinity the entropy diverges as minus
$\ln{\beta}$ since $S_{B}$ is zero in that limit. The problem is not
that the entropy is negative, but the fact that it is unbounded.
$\beta$-duality leads also to an intriguing relationship between the
Helmholtz free energy and the average of the entropy, namely,
\begin{equation}
  -\beta F(\beta,L)= \frac{1}{2}\left[S(\beta,L)+
  S\left(\frac{4\pi^{2}\alpha^{'}}{\beta},L \right)\right]\;.
\end{equation}
Another consequence of $\beta$-duality is that the energy density
$\rho(\beta,L)$ vanishes at the self-dual temperature $\beta=2 \pi
\sqrt{\alpha^{'}}$ for every value of the volume. Similarly,
$L$-duality implies the vanishing of the pressure for $L=2
\pi\sqrt{\alpha^{'}}$ and any value of $\beta$. This means that at
the
self-dual temperature and the self-dual size we have $\rho=p=0$. We
will see in section 3 that this particular situation would define a
static euclidean universe.

The main question is that of the existence of any connection between
$F(\beta)$ given by (\ref{free-energy}) and any quantum field theory.
As mentioned in \cite{OV-M-1}, (\ref{free-energy}) does not coincide
with the free energy of a massless scalar boson. Finding a
relationship with fields is equivalent to distinguishing a stringy
regime. The presence of $\alpha^{'}$ induces us to identify the
field-theoretical regime as that arising in the limit $\beta L \gg 4
\pi^{2}\alpha^{'}$, to get
\begin{equation}
  F(\beta,L)\rightarrow -\frac{L}{24 \pi \alpha^{'}}
  +\frac{1}{\beta}\ln{\left(\frac{\beta}{\sqrt{\alpha^{'}}}\right)}
  +\frac{2}{\beta}\ln{\eta\left(i\frac{\beta}{L}\right)}\;,
\end{equation}
since in this approximation we have that
\begin{equation}
  -\frac{1}{\beta}\left[\frac{\beta L}{24 \pi \alpha^{'}}
  +\ln{\left(\frac{\beta}{\sqrt{\alpha^{'}}}\right)}\right] \sim
  -\frac{L}{24 \pi \alpha^{'}}\;,
\label{lim1}
\end{equation}

we finally obtain the free energy for a massless boson in $S^{1}
\times \mbox{\bf R}$ with an additional vacuum energy given by
$-L/(24
\pi \alpha^{'})$. The corresponding equation of state is
$p=\rho+L/(12
\pi \alpha^{'})$.

On the other hand we can also identify an {\it ultrastringy} regime
in
which $\beta L \ll 4 \pi^{2}\alpha^{'}$. In this limit the Helmholtz
free energy takes the form
\begin{equation}
  F(\beta,L)\rightarrow -\frac{2\pi^{3}\alpha^{'}}{3\beta^{2}L}+
  \frac{2}{\beta}\ln{\eta\left(i\frac{L}{\beta}\right)}
\end{equation}
which gives rise to the equation of state $p=\rho-4
\pi^{3}\alpha^{'}/(3\beta^{2}L^{2})$. So, in this regime, the string
cannot be described in terms of a quantum field theory, as it must be
if string theory is something more than ordinary quantum field
theory.

We have seen that the free energy (\ref{free-energy}) suffers from a
severe problem, namely, the canonical entropy diverges when $\beta$
goes to infinity. This is a consequence of the presence of the term
\begin{equation}
  \frac{1}{\beta}\ln{\left(\frac{\beta}{\sqrt{\alpha^{'}}}\right)}
\end{equation}
in the free energy which ensures $\beta$-duality and the invariance
of
$\beta F(\beta,L)$ under the replacement $\beta \leftrightarrow L$.
In
order to solve the problems with (\ref{free-energy}) we propose that
the correct result for the free energy of the $c=1$ string at finite
temperature is obtained by dropping the term containing
$\ln{(\beta/\sqrt{\alpha^{'}})}$ in (\ref{free-energy}), namely
\begin{equation}
  \hat{F}(\beta,L)=F_{B}(\beta,L)+F_{B}\left(\beta,\frac{4\pi^{2}
    \alpha^{'}}{L}\right)=
  \frac{2}{\beta}\ln{\eta\left(i\frac{\beta}{L}\right)}+
  \frac{2}{\beta}\ln{\eta\left(i\frac{\beta L}{4\pi^{2}\alpha^{'}}
  \right)}\;.
\label{our}
\end{equation}
The new equation of state is
\begin{equation}
  \rho-p=-\frac{1}{12\pi\alpha^{'}} E_{2}\left(i\frac{\beta
    L}{4\pi^{2}\alpha^{'}}\right)\;.
\end{equation}
That $\hat{F}$ gives a positive definite canonical entropy is a
trivial exercise since it is the sum of the entropies for two bosonic
fields. What is more important is to check that using (\ref{our}) we
can recover the field-theoretical regime when $\beta L\gg 4
\pi^{2}\alpha^{'}$. Using the expression of the Dedekind
$\eta$-function we get
\begin{equation}
  \hat{F}(\beta,L)\rightarrow -\frac{L}{24\pi\alpha^{'}}+
  \frac{2}{\beta}\ln{\eta\left(i\frac{\beta}{L}\right)}\;,
\end{equation}
i.e., we get the expression for a massless field with a vacuum
energy,
in accordance with the result (\ref{lim1}) obtained from
(\ref{free-energy}). In the {\it ultrastringy} regime ($\beta L \ll 4
\pi \alpha^{'}$) the limit we get has the same form than the one
obtained from (\ref{free-energy}),
\begin{equation}
  \hat{F}(\beta,L)\rightarrow -\frac{2\pi^{3}\alpha^{'}}{3\beta^{2}L}
  +\frac{2}{\beta}\ln{\eta\left(i\frac{\beta}{L}\right)}\;.
\end{equation}
Nevertheless, we have to use an aproximation similar to that in
(\ref{lim1}).

As we have pointed out, $\beta\hat{F}(\beta,L)$ does not present
either $\beta$-duality or invariance under the replacement
$\beta\leftrightarrow L$. Nevertheless, as mentioned in sec. 1,
$\beta$-duality does not survive the coupling to Brans-Dicke gravity
so it seems that this is not a fundamental symmetry to preserve. With
respect to the breaking of the invariance under the interchange of
$\beta$ and $L$ this is a consequence of the fact that the proposed
free energy is not obtained from the toroidal compactification in
$S^{1}\times S^{1}$. We have shown \cite{OV-M-1} that the same
breakdown of the equivalence between the Helmholtz free energy and
the
toroidal compactification happens when studying massless fields
living
in compact spaces, in particular in $S^{1}\times \mbox{\bf R}$. What
we claim is that this is what happens when considering a
two-dimensional string in $S^{1}\times\mbox{\bf R}$: the Helmholtz
free energy is not given by the toroidal compactification. We state
that the free energy of the two-dimensional string can be interpreted
as the one corresponding to two massless fields in
$S^{1}\times\mbox{\bf R}$, one living in a circumference of length
$L$
and the other one in a circumference of length $4\pi^{2}\alpha^{'}/L$
(cf. \cite{OV}).  Notice that (\ref{our}) enjoys invariance under the
duality transformation $L\rightarrow 4\pi^{2}\alpha^{'}/L$. Since
this
is a symmetry of the Brans-Dicke action (with the {\it ansatz}
(\ref{FRW}) for the metric) this seems to be the preserved
fundamental
symmetry.

In the next section we will study the solutions to the
Einstein-Hilbert-Brans-Dicke equations using either
(\ref{free-energy}) or (\ref{our}) for the Helmholtz free energy.

\section{Cosmological solutions}

We are interested in the cosmological solutions of our stringy
universe. Then we are going to consider our string propagating in the
presence of a background metric $g_{\mu\nu}$ and a background dilaton
field $\Phi$. The condition for this background to define a CFT (to
lowest order in $\alpha^{'}$) is given by the vanishing of the
$\beta$-functions associated with the background fields.  The
equations so obtained for $g_{\mu\nu}$ and $\Phi$ can also result
from
using the action principle with the following action functional
\cite{TV}
\begin{equation}
  S_{eff}= \int d^{2}\xi \sqrt{-g}
\left[\Phi(R+\frac{16}{\alpha^{'}})
  + \frac{1}{\Phi}\nabla_{\mu}\Phi\nabla^{\mu}\Phi\right]
\end{equation}
which correspond to the Brans-Dicke action (\ref{EHBD}) with
$\omega=-1$ and $\Lambda=-8/\alpha^{'}$. The coupling to the matter
can be made via the perfect-fluid energy-momentum tensor
\begin{equation}
  T_{\mu\nu}=(p+\rho)u_{\mu}u_{\nu}-pg_{\mu\nu}
\end{equation}
The equations for the background fields are then
\begin{eqnarray}
  -\frac{8}{\alpha^{'}}g_{\mu\nu}&=&\frac{8\pi}{\Phi}T_{\mu\nu}-
  \frac{1}{\Phi^{2}}
  \left(\nabla_{\mu}\Phi\nabla_{\nu}\Phi-\frac{1}{2}g_{\mu\nu}
  \nabla_{\sigma}\Phi\nabla^{\sigma}\Phi\right)\nonumber \\ &+&
  \frac{1}{\Phi}\left(\nabla_{\mu}\nabla_{\nu}\Phi
  -g_{\mu\nu}\Box\Phi\right)\;,
\label{eqn1}\\
R+\frac{16}{\alpha^{'}}&=&
-\frac{1}{\Phi^{2}}g^{\mu\nu}\nabla_{\mu}\Phi\nabla_{\nu}\Phi+
\frac{2}{\Phi}\Box\Phi\;,
\label{eqn2}
\end{eqnarray}
together with the integrability condition imposed by the local
conservation of the energy-momentum tensor $\nabla_{\mu}
T^{\mu\nu}=0$.

For the background metric we are going to use the
Friedmann-Roberston-Walker {\it ansatz} (\ref{FRW}) and a
space-independent dilaton field $\Phi(t)$. In this case eqs.
(\ref{eqn1}) and (\ref{eqn2}) can be rewritten as
\begin{eqnarray}
  2\Phi^{2}\frac{\ddot{L}}{L}+\frac{16}{\alpha^{'}}\Phi^{2}
  &=&\dot{\Phi}^{2}-2
  \Phi\ddot{\Phi}-2\Phi\dot{\Phi}\frac{\dot{L}}{L}\;,
\label{BD1} \\
\dot{\Phi}^{2}+2\Phi\dot{\Phi}\frac{\dot{L}}{L} &=&
16\pi\Phi\rho(\beta,L)-\frac{16}{\alpha^{'}}\Phi^{2}\;,
\label{BD2} \\
\Phi\ddot{\Phi}-\frac{1}{2}\dot{\Phi}^{2} &=& -8\pi \Phi
p(\beta,L)-\frac{8}{\alpha^{'}}\Phi^{2}\;, \label{1.3}
\end{eqnarray}
and the energy-momentum conservation
\begin{equation}
  \dot{\rho}(\beta,L)+\frac{\dot{L}}{L}[\rho(\beta,L)+p(\beta,L)]=0.
\label{tmunu}
\end{equation}
It is easy to check that eq. (\ref{BD1}) is equivalent to the
conservation of $T_{\mu\nu}$ so we can drop it and be left with eqs.
(\ref{BD2}), (\ref{1.3}) and (\ref{tmunu}) that together with the
equation of state determine our system in terms of $\Phi(t)$, $L(t)$
and $\beta(t)$.

Let us now make some comments about $L$-duality. We have seen that
$\hat{F}(\beta,L)$ (and $F(\beta,L)$) are invariant under the
replacement $L\rightarrow 4\pi^{2}\alpha^{'}/L$. This implies that
the
energy density and the pressure transform according to
\begin{equation}
  \rho(\beta,L)=\frac{4\pi^{2}\alpha^{'}}{L^{2}} \rho \left(\beta,
  \frac{4\pi^{2}\alpha^{'}}{L}\right)\;, \hspace{2cm}
  p(\beta,L)=-\frac{4\pi^{2}\alpha^{'}}{L^{2}}p
  \left(\beta,\frac{4\pi^{2}\alpha^{'}}{L}\right)\;.
\end{equation}
It can be easily checked that these transformations for the energy
density and the pressure, together with the transformation of the
dilaton field $\Phi(t)$ make the system of differential equations
(\ref{BD2}), (\ref{1.3}) and (\ref{tmunu}) invariant under
$L$-duality
(in fact, the equations combine among themselves). A pending problem
is that of clarifying what is the dynamical meaning of this symmetry.
It has been argued in \cite{BV,TV} that the duality property would
imply an effective minimum length $L_{min}=2\pi\sqrt{\alpha^{'}}$ for
the universe. Nevertheless, in \cite{BV}, the arguments leading to
this conclusion lie upon the condition that one has to substitute the
definition of the localized states in terms of the Fourier transform
of the momentum modes when $L>L_{min}$ (which are light in this
regime), by the definition in terms of the Fourier transform of the
winding modes when $L$ decreases below the self-dual length (which
are
the light states in this case). Nevertheless, the only conclusion one
can extract from $L$-duality using equations
(\ref{BD2})-(\ref{tmunu})
is that, for any solution $L(t)$, $\Phi(t)$ and $\beta(t)$, the new
set of functions
\begin{equation}
  L^{'}(t)=\frac{4\pi^{2}\alpha^{'}}{L(t)}, \hspace{1cm}
  \Phi^{'}(t)=\frac{L^{2}(t)}{4\pi^{2}\alpha^{'}}\Phi(t)\;,
  \hspace{1cm} \beta^{'}(t)=\beta(t)\;,
\label{dual}
\end{equation}
is also a solution to (\ref{BD2})-(\ref{tmunu}). In fact, as we will
see in a moment, there are solutions for which $L(t)$ goes through
the
self-dual length and ends at $L=0$. This kind of solutions exist when
we take $F(\beta,L)$ as the free energy as well as when
$\hat{F}(\beta,L)$ is chosen.

With respect to $\beta$-duality, the situation is somewhat different.
The transformation property of $F(\beta,L)$ with respect to the
duality replacement $\beta\rightarrow 4\pi^{2}\alpha^{'}/\beta$
implies the following transformation rules for the energy density and
the pressure
\begin{equation}
  \rho(\beta,L)=-\frac{4\pi^{2}\alpha^{'}}{\beta^{2}}\rho \left(
  \frac{4\pi^{2}\alpha^{'}}{\beta},L\right)\;, \hspace{1cm}
  p(\beta,L)=\frac{4\pi^{2}\alpha^{'}}{\beta^{2}} p\left(
  \frac{4\pi^{2}\alpha^{'}}{\beta},L\right)\;.
\end{equation}
It is now clear from eqs. (\ref{BD2})-(\ref{tmunu}) that
$\beta$-duality is broken by the coupling the two-dimensional
dilatonic gravity. This is because, unless the case of $L$-duality,
the geometric part of the action, i.e., $L(t)$ and $\Phi(t)$, do not
transform under $\beta$-duality, so one cannot reabsorb the change
induced in the energy density and the pressure by a change in the
dynamical variables $L$ and $\Phi$.

When solving the classical equations of motion for the fields, there
are a set of variables in terms of which the expressions simplify
notably and which have a clear physical meaning. Let us define two
new
functions $u(t)$, $v(t)$ in terms of $\beta(t)$ and $L(t)$
\begin{equation}
  u(t)=\frac{L(t)}{\beta(t)}\;, \hspace{1cm} v(t)=\frac{\beta(t)
    L(t)}{4\pi^{2}\alpha^{'}}\;.
\end{equation}
Here $v$ is a kind of {\it stringiness} parameter; when $v\gg 1$ we
are in what we called before the field-theoretical regime. In that
case, since the string Helmholtz free energy (either $F$ or
$\hat{F}$)
reduces to the free energy for a massless boson in
$S^{1}\times\mbox{\bf R}$ plus a vacuum energy, we expect the
universe
to behave as in the cases studied in \cite{OV-M-1} with the
appropriate values for $\omega$ and $\Lambda$.  On the other hand,
the
limit $v\ll 1$ corresponds to a {\it ultrastringy} regime in which
the
matter content of the universe is governed by the equation of state
deduced in sec. 2.

\begin{figure}[h]
  \epsffile{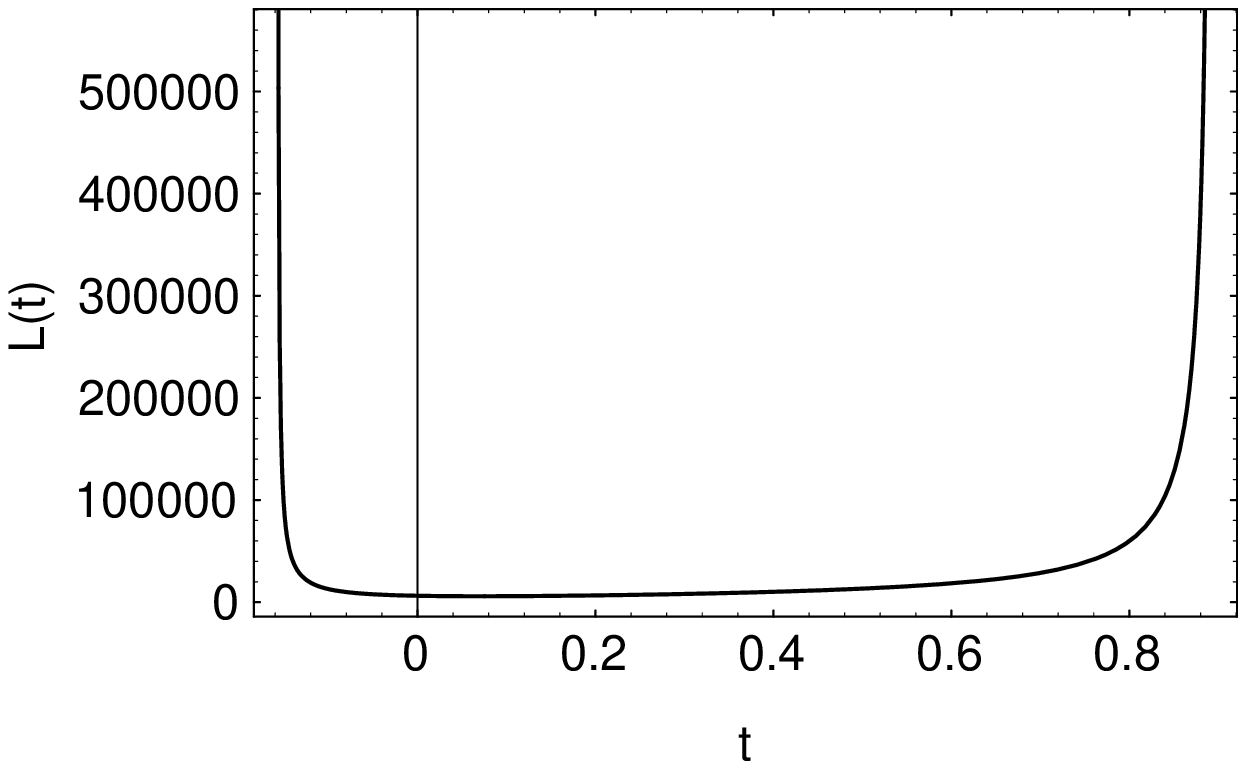}
\caption{Scale factor $L$ vs. $t$ for $v\gg 1$.}
\label{FieldL}
\end{figure}

\begin{figure}[h]
  \epsffile{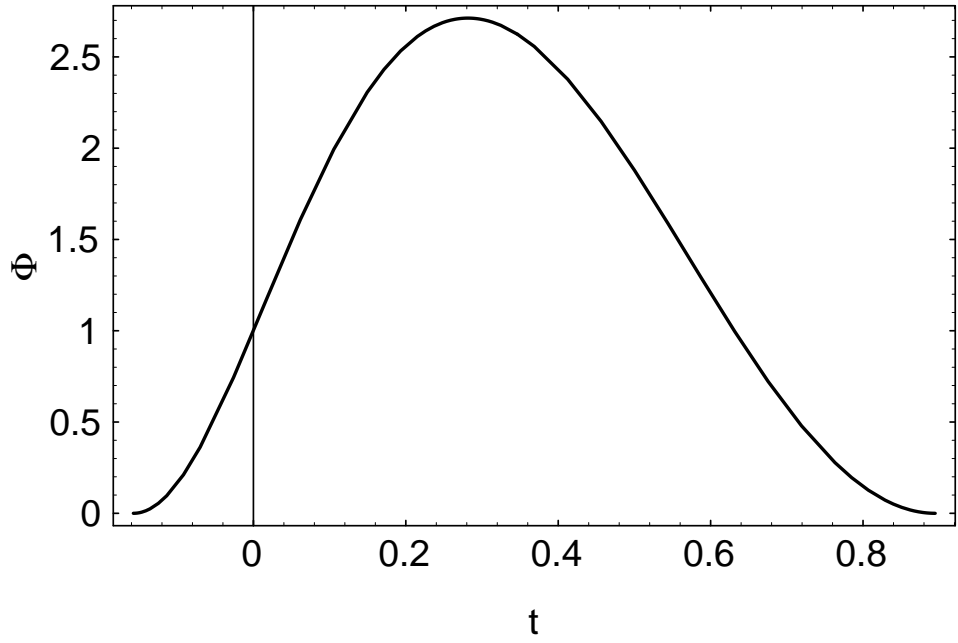}
\caption{Dilaton vs. $t$ for $v\gg 1$.}
\label{FieldD}
\end{figure}

\begin{figure}[h]
  \epsffile{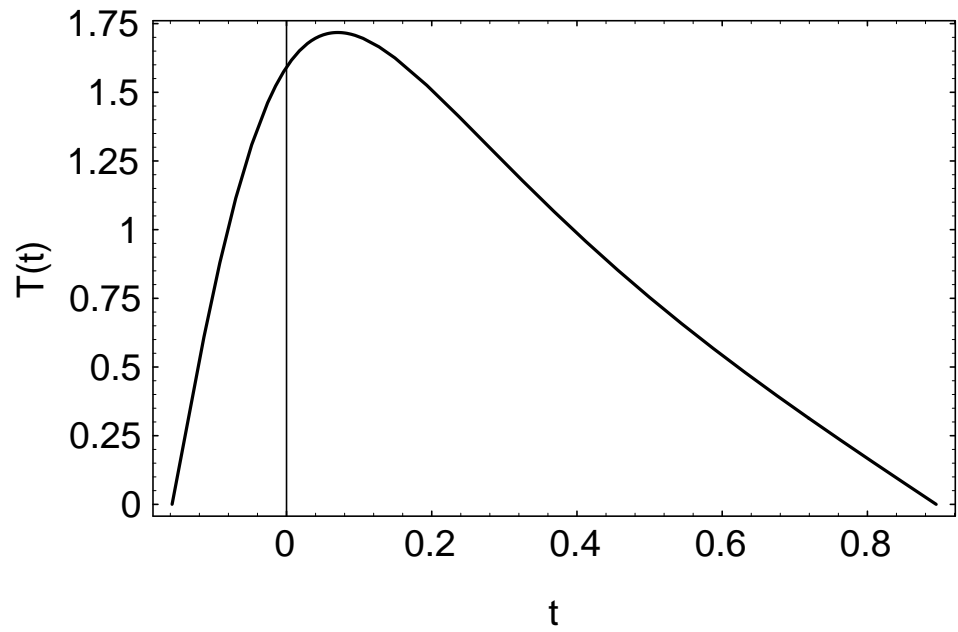}
\caption{Temperature vs. $t$ for $v\gg 1$.}
\label{FieldT}
\end{figure}

When studying the numerical solutions to (\ref{BD2})-(\ref{tmunu})
using the matter described by (\ref{our}) one can find three
different
kinds of solutions depending on the initial conditions $u_{0}$,
$v_{0}$, $\Phi_{0}$ and $\dot{\Phi}_{0}$. The first one corresponds
to
field-theoretical-like universes in which $v(t)\gg 1$ for every value
of $t$. In fig. \ref{FieldL} we have plotted $L(t)$ for this class of
solutions. The universe contracts from an infinite size and reaches a
minimum length that, however, is much larger than the self-dual size.
The dilaton field (fig. \ref{FieldD}) grows and, after reaching a
maximum, begins to decrease. In fig. \ref{FieldT} we have plotted the
time evolution of the temperature for this universe. We see that, as
it corresponds to a field-theoretical universe, the temperature drops
to zero when the size of the universe goes to infinity (cf. the
results gotten in ref. \cite{OV-M-1}).

\begin{figure}[h]
  \epsffile{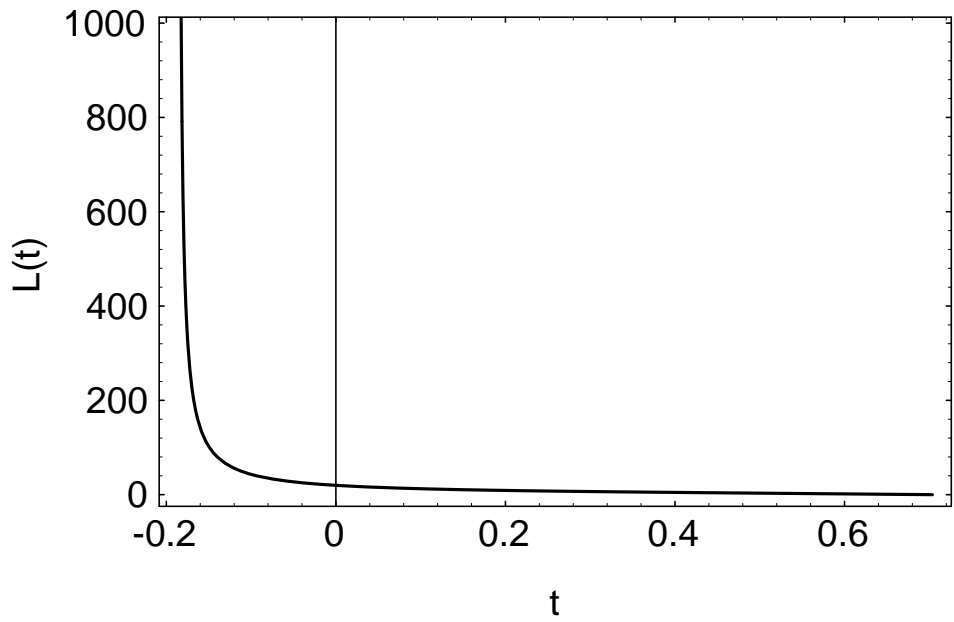}
\caption{Scale factor vs. $t$.}
\label{twoL}
\end{figure}

\begin{figure}[h]
  \epsffile{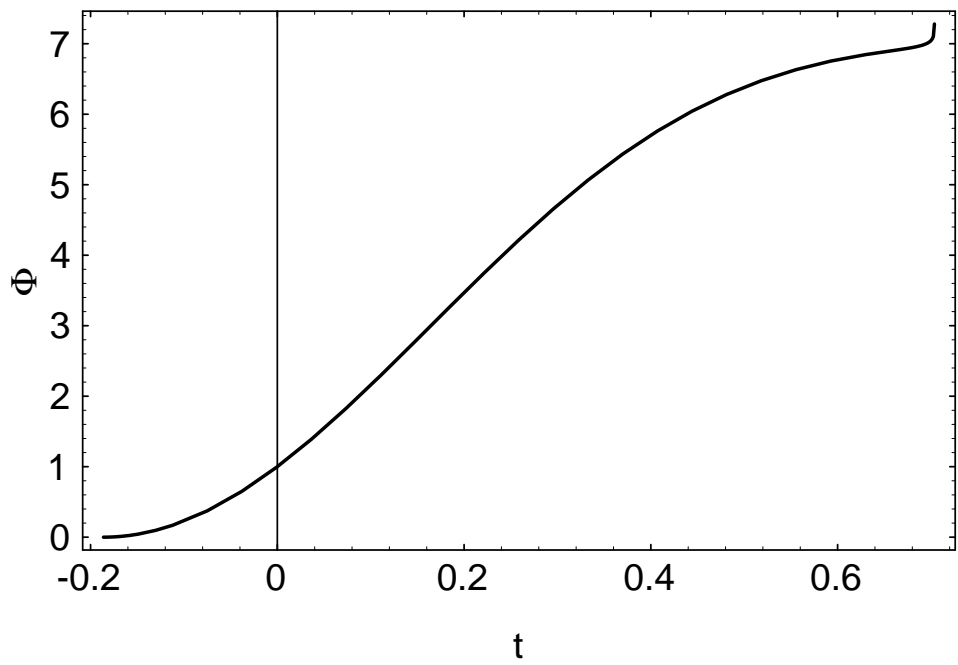}
\caption{Dilaton field vs. $t$.}
\label{twoD}
\end{figure}

\begin{figure}[h]
  \epsffile{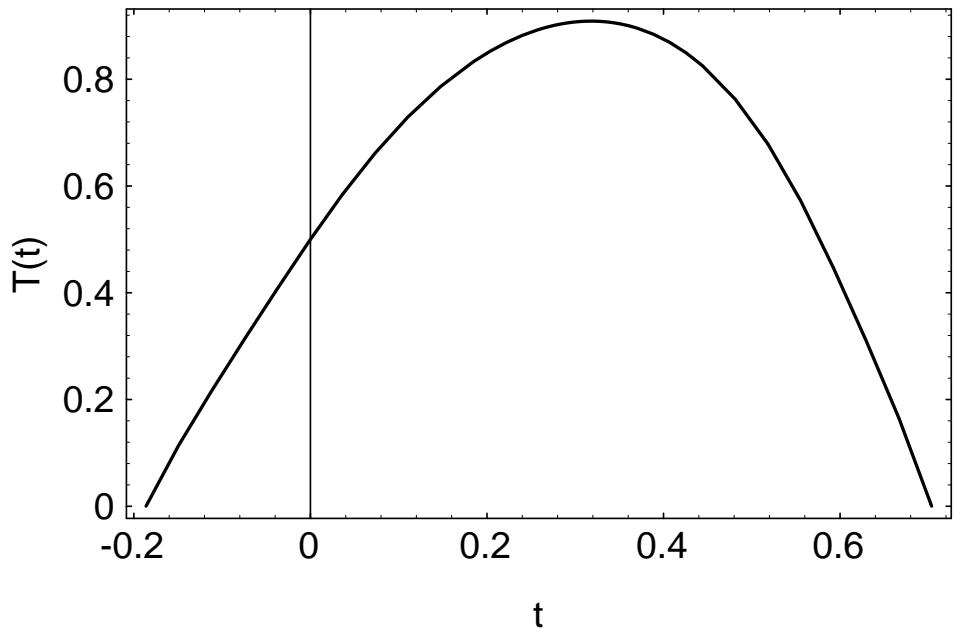}
\caption{Temperature vs. $t$.}
\label{twoT}
\end{figure}

A second kind of solutions arise when we consider less restrictive
initial conditions. In this case we get a universe whose scale factor
$L(t)$ comes from infinity and after reaching the self-dual length
keeps on decreasing until it gets $L=0$ (fig. \ref{twoL}). The
dilaton, on the other hand, grows monotonously as it is shown in fig.
\ref{twoD}. The temperature, nevertheless, has a quite stringy
behavior (fig. \ref{twoT}). For small $t$, when the universe is big,
the temperature is low, which is consistent with a field-theoretical
description. However, when the size of the universe gets below the
self-dual size, the temperature, instead of continuing raising (as
one
expects from field theory) begins to decrease and it is zero when the
universe reaches a null size. This behavior is a consequence of
$L$-duality, because the entropy satisfies
$S(\beta,L)=S(\beta,4\pi^{2}\alpha^{'}/L)$. Since the integrability
condition $\nabla_{\mu}T^{\mu\nu}=0$ implies that the entropy is
conserved, we have
\begin{equation}
  \frac{d}{dt}S(\beta,L)=0=\frac{\partial S}{\partial
    \beta}\dot{\beta} +\frac{\partial S}{\partial L}\dot{L}\;.
\end{equation}
Being the scale factor $L(t)$ a monotonously decreasing function, we
can parametrize the evolution of $\beta$ by $L$. Then, from the last
relation we have
\begin{equation}
  \frac{d\beta}{dL}=\frac{\dot{\beta}}{\dot{L}}= -\frac{\partial
    S/\partial L}{\partial S/\partial \beta}[\beta(L),L]\;.
\end{equation}
Now, implementing duality it is easy to check that
\begin{equation}
  \frac{d\beta}{dL}[\beta(L),L]\frac{d\beta}{dL}
  \left[\beta\left(\frac{4\pi^{2}\alpha^{'}}{L}\right),
  \frac{4\pi^{2}\alpha^{'}}{L}\right] <0\;,
\end{equation}
so this implies that $d\beta/dL=0$ if $L=2\pi\sqrt{\alpha^{'}}$.

\begin{figure}[h]
  \epsffile{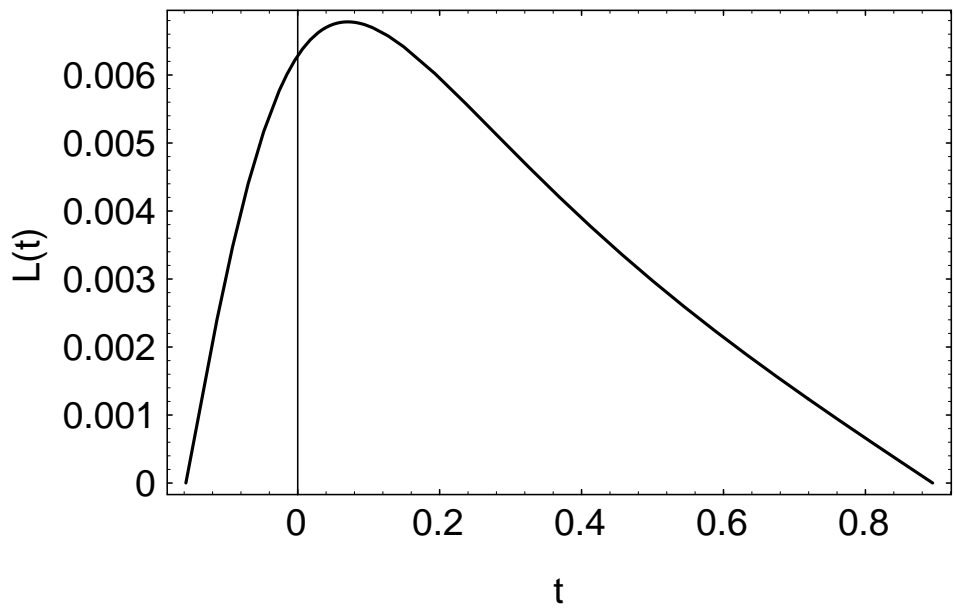} \caption{Scale factor vs. $t$ for the {\it
      ultrastringy} kind of solutions.} \label{usL} \end{figure}

\begin{figure}[h]
  \epsffile{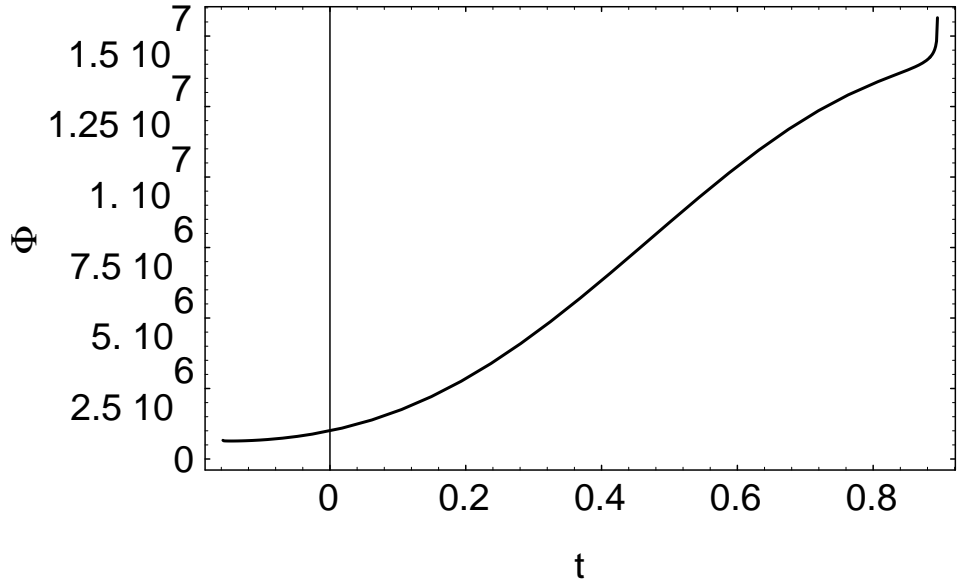} \caption{Dilaton field vs. $t$ for the {\it
      ultrastringy} kind of solutions.} \label{usD} \end{figure}

We have said above that the meaning of $L$-duality is that the
solutions can be paired according to (\ref{dual}). When we apply the
duality transformations to the second kind of solutions describing a
collapsing universe, we get new solutions which closely resembles the
time reversal of the original ones, i.e., we have a scale factor that
starts at $L=0$ and grows monotonously, passing through the self-dual
value, and reaching infinity in a finite time. Nevertheless, when we
apply (\ref{dual}) to the first kind of solutions (the
field-theoretical-like ones) we get a new class featuring a string
that lives in the {\it ultrastringy} regime all the time. The scale
factor $L(t)$ starts and ends at zero length and reaches a maximum
value, much smaller than the self-dual one (fig. \ref{usL}). With
respect to the dilaton field we get a monotonously increasing
$\Phi(t)$ very different from the field-theoretical type (fig.
\ref{usD}). The temperature, on the contrary, is the corresponding to
the field-theoretical-like case (fig. \ref{FieldT}).

One can ask oneself whether there is a solution describing a static
universe. Imposing the condition $\dot{L}=\ddot{L}=0$ on eqs.
(\ref{BD2})-(\ref{tmunu}) we find that there is one given by
\begin{equation}
  L(t)=2\pi\sqrt{\alpha^{'}}, \hspace{1cm} \beta(t) \sim
  2\pi\sqrt{\alpha^{'}} \times 0.523522\;, \hspace{1cm}
  \Phi(t)=C\exp{\left(i \frac{4t}{\sqrt{\alpha^{'}}}\right)}\;,
\end{equation}
where the values of $L$ and $\beta$ are fixed by the requirement that
the energy density and the pressure vanish. We dislike complex
dilatonic fields since the $\sigma$-model dilaton is a real field.
This can be fixed by performing a Wick rotation to imaginary time
$t\rightarrow it$. In that case we get a solution to Euclidean
dilatonic gravity with constant scale factor. This solution can be
interpreted as a kind of gravitational instanton interpolating
between
two universes, both at the self-dual size, one of which is in a
strong
coupling regime ($\Phi\ll 1$) and the other one in a weak coupling
one
($\Phi\gg 1$) since we relate the string coupling constant with the
vacuum expectation value of the dilaton field. Besides, as it is the
case with most of the known gravitational instantons, this
configuration of the fields renders the geometric part of the action
zero. One can see that the static situation results from the action
of
the negative Casimir energy which compensates the positive
contribution of the thermal motion to the internal energy.

\begin{figure}[h]
  \epsffile{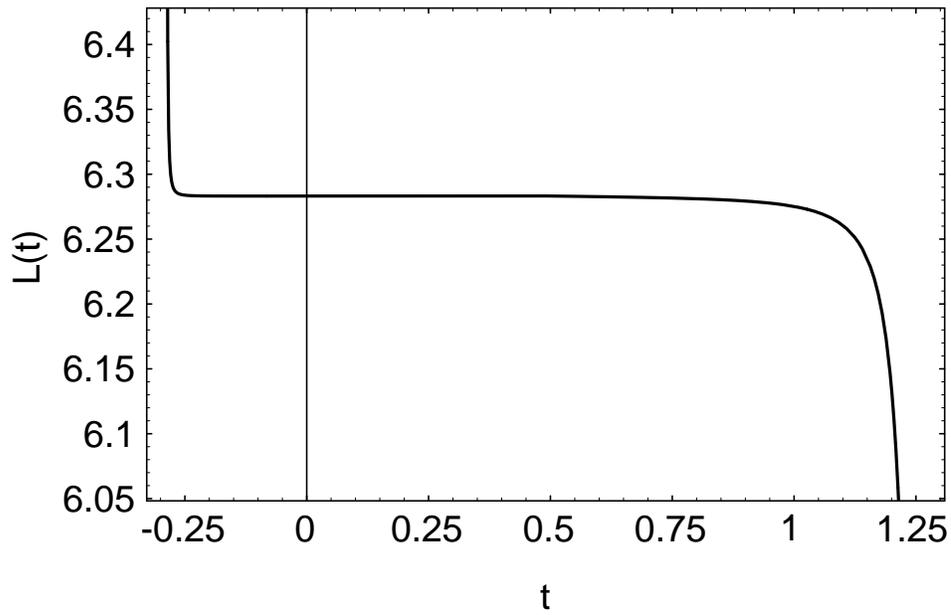}
\caption{Scale factor vs. $t$ for the {\it intermediate} case.}
\label{intL}
\end{figure}

\begin{figure}[h]
  \epsffile{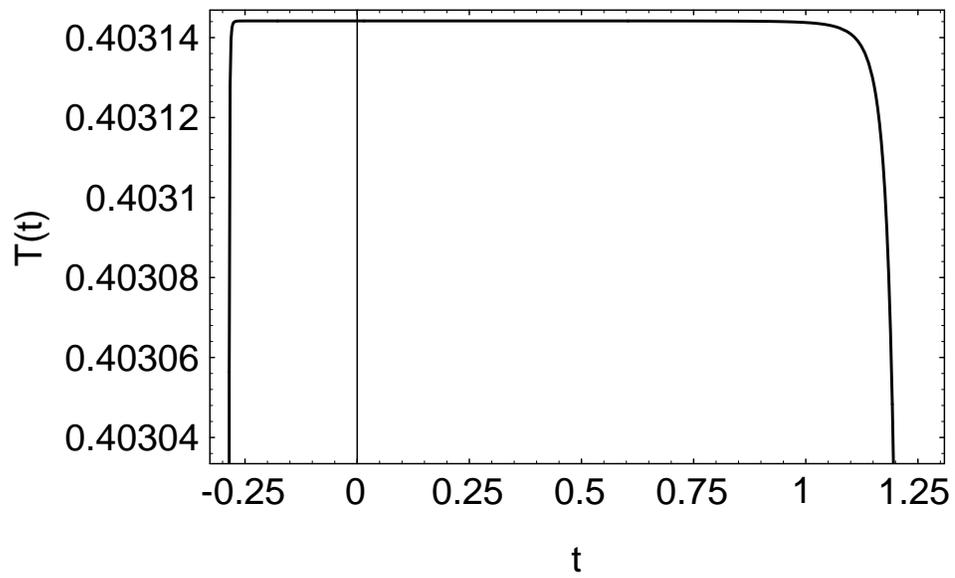} \caption{Temperature vs. $t$ for the {\it
      intermediate} case.} \label{inT} \end{figure}

Now we are going to study how all this structure of solutions changes
when we substitute the Helmholtz free energy given by (\ref{our}) by
the one gotten from the toroidal compactification, i.e.
(\ref{free-energy}). The first thing we can see is that, since in
this
case $F(\beta,L)$ recovers the field theory regime when $v\gg 1$, we
are operatoralso going to have field-theoretical-like solutions of
the
type described above. On the other hand the {\it ultrastringy}
solutions gotten from them by performing the duality transformation
(\ref{dual}) are also to be present (let us recall that $F$ does
enjoy
$L$-duality as well as $\beta$-duality). An important point
nevertheless is that now we have a kind of {\it intermediate}
solutions of the type depicted in figs. \ref{intL} and \ref{inT}.
This intermediate (marginal) case sitting between the first
(field-theoretical) and the second types can be gotten by making a
fine-tuning of the initial conditions. The corresponding universe has
a scale factor (see fig. \ref{intL}) which comes from infinity and,
after reaching the self-dual length, it stays at that value of $L$
for
a long time and finally drops to $L=0$. Of course, by applying
$L$-duality to (\ref{dual}) we get a similar exploding solution
starting at $L=0$. The temperature now follows a curve similar to
that
of fig. \ref{twoT}, i.e., it is zero in both limits, but with a
pronounced plateau (fig. \ref{inT}). Besides, we also find those
solutions describing a monotonously contracting (resp. expanding)
universe ending (resp. beginning) at $L=0$ of the type described by
figs. \ref{twoL}-\ref{twoT}. The instanton-like solution described
above, is also present but now the existence of $\beta$-duality sets
the temperature to the self-dual value.

In both cases (for $\hat{F}$ and $F$) the field equations are
invariant under time reversal. This means that, together with the
type
of solutions described above we also have their time reversed ones
which differ by the sign of $\dot{\Phi}_{0}$. All the plots have been
made in units in which $\alpha^{'}=1$.

\section{Conclusions}

We have studied the cosmological solutions to dilatonic gravity using
as the source of the gravitational field a gas of two-dimensional
strings. The first thing we have found is that the expression of the
Helmholtz free energy given by (\ref{free-energy}) and deduced in
\cite{DKL} presents an important problem; it gives a canonical
entropy
which is negative and diverges in the low temperature limit. As a
solution we have proposed for the Helmholtz free energy another
expression, (\ref{our}), in which we drop the term which produces the
problem. The principal feature of this new expression is that it
breaks $\beta$-duality and, consequently, invariance under the
interchange $\beta\leftrightarrow L$. The new expression can be
interpreted as the free energy for two massless bosons one of which
lives in a circumference of length $L$ and the other one in a
circumference of length $4\pi^{2}\alpha{'}/L$ (cf. \cite{OV}). This
breaking of the equivalence between the toroidal compactification and
the Helmholtz free energy can be traced back to the massless bosonic
field in $S^{1}\times \mbox{\bf R}$ \cite{OV-M-1}. Since the only
on-shell state of the $c=1$ string (despite of the topological
discrete states, which do not contribute to the partition function
\cite{KP}) is the center of mass which can be effectively described
by
a massless bosonic field \cite{P}, it seems to be natural to keep
this
equivalence at finite temperature. Indeed, the effective image of the
$c=1$ string (or equivalently, the critical two-dimensional one) is
that of its center of mass. If we look at it as a point on a
cylinder,
its hamiltonian would be given by $\hat{H}_{1}=\hat{m}/L$ where
$\hat{m}$ is the operator whose eigenvalues are the momentum integer
numbers of the string. On the other hand, since the string is an
extended object it can wrap around the spatial compactified dimension
and we have another contribution to its energy given by
$\hat{H}_{2}=\hat{n}L/(4\pi^{2}\alpha^{'})$, $\hat{n}$ being the
winding number operator with integer eigenvalues $n$. Then, from a
target-space point of view the dynamics of the string is governed by
the hamiltonian
\begin{equation}
  \hat{H}=\frac{\hat{m}}{L}+\frac{\hat{n}L}{4\pi^{2}\alpha^{'}}
\end{equation}
To compute the thermal partition function of the multi-string system
we need to make a trace over the second quantized states
$|m_{i},n_{i}\rangle$ which are eigenvectors of the operators
$\hat{m}_{i}$ and $\hat{n}_{i}$. Since the complete Hilbert space is
the direct product of the winding and the momentum sectors the
partition function factorizes and this is reflected in the fact that
the Helmholtz free energy is the sum of two terms which correspond to
two massless fields one in a space with length $L$ and the other one
in the dual space. Now it is important to notice that the zero mode
in
both traces has to be suppressed. By the way, the partition function
in the zero temperature limit computed in \cite{P} can be interpreted
as the partition function of two bosonic massless fields one at a
temperature $T=1/L$ and the second one at $T=L/(4\pi^{2}\alpha^{'})$

After studying the cosmological solutions, using both $F$ and
$\hat{F}$ we arrive at several conclusions. First of all, we find
that
there is no dynamical rebound of the scale factor when reaching the
self-dual size. Secondly we find that our string universe is not free
from singularities. In fact, we have solutions that begin or end at
$L=0$. At first sight this is quite surprising because they are
absent
in a field-theoretical universe (see \cite{OV-M-1}). But one of the
claims in \cite{BV} is that windings prevent expansion and that is
indeed what they do so as to get $L=0$. What we do not see is the
suspected behaviour of the momentum states preventing the
contraction.
This effect would be the result of changing from the momentum modes
when $L>2\pi\sqrt{\alpha^{'}}$ to the winding modes when
$L>2\pi\sqrt{\alpha^{'}}$. In some sense we can say that the
existence
of this kind of singular solutions is implied by duality since the
equivalence of $L=0$ and $L=\infty$ is a result of this symmetry.

It is really important to notice that when all the spatial dimensions
are compact the thermal free energy is not given by the corresponding
toroidal compactification if massless fields are involved. For
example, this means that the free energy for the heterotic
supersymmetric string with all the spatial dimensions compactified is
not given by the toroidal compactification because the massless
excitations of the string has to be treated with care. At least this
is pretty clear from the analogue model in which we compute the free
energy summing over the field content. Using this approach in a
proper
time representation of the free energy we know that the massless
sector is not going to be correctly described. On the contrary, the
modular invariant result for the associated toroidal compactification
(see second reference in \cite{OV-M2}) does not present any
divergence
for any value of $\beta$ when $Im\,\tau$ goes to infinity where
$\tau$
is the standard one-loop modular parameter. What happens is that the
connection between the modular invariant result and the proper time
implementation of the analogue model is broken for any value of
$\beta$. If we forget about this way of representing the analogue
model we can try to implement it by directly writting the
contribution
of the massless sector as computed in \cite{OV-M-1} and then adding
the contribution of the massive states in the proper time
representation which is perfectly well defined. It seems that now the
question about modular invariance is nonsense.

As a final comment, the qualitative effect of substituting
$\hat{F}(\beta,L)$ by $F(\beta,L)$ is only to add the pseudo-statical
solutions of fig. \ref{intL}; we preserve the other three types of
solutions discussed in sec. 3.

\section*{Acknowledgements}

We thank J.\,L.\,F. Barb\'on for valuable discussions. We acknowledge
finantial support by the CICyT Project No. AEN-90-0272. The work of
M.A.V.-M. has also been partially supported by a Comunidad Aut\'onoma
de Madrid Predoctoral Fellowship.

\newpage

 \end{document}